# Coherent spin rotation-induced zero thermal expansion in MnCoSi-based spiral magnets

Jun Liu[1], Bei Ding[1], Yuan Yao[1], Xuekui Xi[1], Zhenxiang Cheng[2], Jianli Wang[2], Chin-wei Wang[3], Guangheng Wu[1] and Wenhong Wang[1,4*]

[1]*Beijing National Laboratory for Condensed Matter Physics, Institute of Physics, Chinese Academy of Sciences, Beijing 100190, China*

[2]*Institute for Superconducting and Electronic Materials, Innovation Campus, University of Wollongong, Squires Way, North Wollongong, New South Wales 2500, Australia*

[3] *Neutron Group, National Synchrotron Radiation Research Center, Hsinchu 30076, Taiwan*

[4] *Songshan Lake Materials Laboratory, Dongguan, Guangdong 523808, China*

*Corresponding Author: E-mail: Wenhong.Wang@iphy.ac.cn. Tel: +86-010-82649247

**Abstract**

Materials exhibiting zero thermal expansion (ZTE), namely, volume invariance during temperature change, can resist thermal shock and are highly desired in modern industries as high-precision components. However, pure ZTE materials are rare, especially those that are metallic. Here, we report the discovery of a pure metallic ZTE material: an orthorhombic $Mn_{1-x}Ni_xCoSi$ spiral magnet. The introduction of Ni can efficiently enhance the ferromagnetic exchange interaction and construct the transition from a spiral magnetic state to a ferromagnetic-like state in MnCoSi-based alloys. Systematic in situ neutron powder diffraction revealed a new cycloidal spiral magnetic structure in *bc* plane at ground state which would transform to the helical spiral in the *ab* plane with increasing temperature. Combined with Lorentz transmission electron microscopy techniques, the cycloidal and helical spin order coherently rotated at varying periods along the *c* axis during the magnetic transition. This spin rotation drove the continuous movement of the coupled crystalline lattice and induced a large negative thermal expansion along *the a* axis, eventually leading to a wide-temperature ZTE effect. Our work not only introduces a new ZTE alloy but also presents a new mechanism by which to discover or design ZTE magnets.

**Introduction**

It is well known that the inherent anharmonicity of phonon vibrations triggers the volume expansion of most solids upon heating. However, when the crystalline lattice couples with ferroelectricity, magnetism and charge transfer, anomalous behaviour during large temperature fluctuations, namely, negative thermal expansion (NTE) behaviour, may be realized.[1-4] NTE materials combined with positive thermal expansion (PTE) materials can reduce the overall coefficient of thermal expansion (CTE) and lead to overall zero thermal expansion (ZTE) composites, which are of great importance in industrial applications as structural components, electronic devices and high-precision instruments.[1,5-7] Unfortunately, such ZTE composite materials have high internal stresses that can cause microcracking during thermal cycling, which significantly diminishes their mechanical performance and lifetime. This problem can be overcome if ZTE materials are single-phase ones with homogeneous internal structure, especially in metallic form. In addition to Invar alloys,[8] a small number of such single-phase ZTE alloys or compounds have been discovered so far, such as $Mn_{1-x}Co_xB$,[9] YbGaGe,[10] $LaFe_{10.6}Si_{2.4}$,[11] $Ho_2Fe_{16}Cr$,[12] $ErFe_{10}V_{1.6}Mo_{0.4}$[13] and $RFe_2$-based compounds.[14-16]

The most found of these materials are magnetic alloys, for which the ZTE behaviour is intimately associated with spontaneous magnetic ordering, known as the magnetovolume effect (MVE). The MVE originating from spin-lattice coupling may weaken or even compensate for anharmonic lattice variations and lead to abnormal thermal expansion behaviour near the magnetic ordering temperature.[17,18] Here, we present a new class of ZTE materials, MnCoSi-based metamagnetic alloys, consisting a homogeneous phase. The ternary equiatomic MnCoSi alloy crystallizes with an orthorhombic structure from the honeycomb hexagonal structure after experiencing a martensitic transition at high temperature (Fig. 1a and Fig. S1). The shortest Co-Si bonds yield wrinkled eight-membered rings, in which Mn-Mn zig-zag chains are embedded. The interconnected Co-Si contacts along the [100] direction form the basic rigid skeleton. Due to the critical nearest Mn-Mn separation, MnCoSi alloy possesses a ground state of nonlinear antiferromagnetism (AFM) and a magnetic-field-induced magnetoelastic transition,[19,20] during which the large inverse magnetocaloric effect and giant magnetostrictive effect is realized.[21-23] In this study, we find that the helical magnetic structure of MnCoSi, as is widely recognized, would transit to a cycloidal spiral AFM at low temperature. Moreover, we report that when the MnCoSi system is tuned by minimal Ni introduction, an ultralow CTE over a wide temperature range can be achieved. The combination of X-ray diffraction (XRD), neutron powder diffraction (NPD) and Lorentz transmission electron microscopy (TEM) techniques reveals a new mechanism underlying ZTE: coherent spin rotation of the spiral magnetic structures.

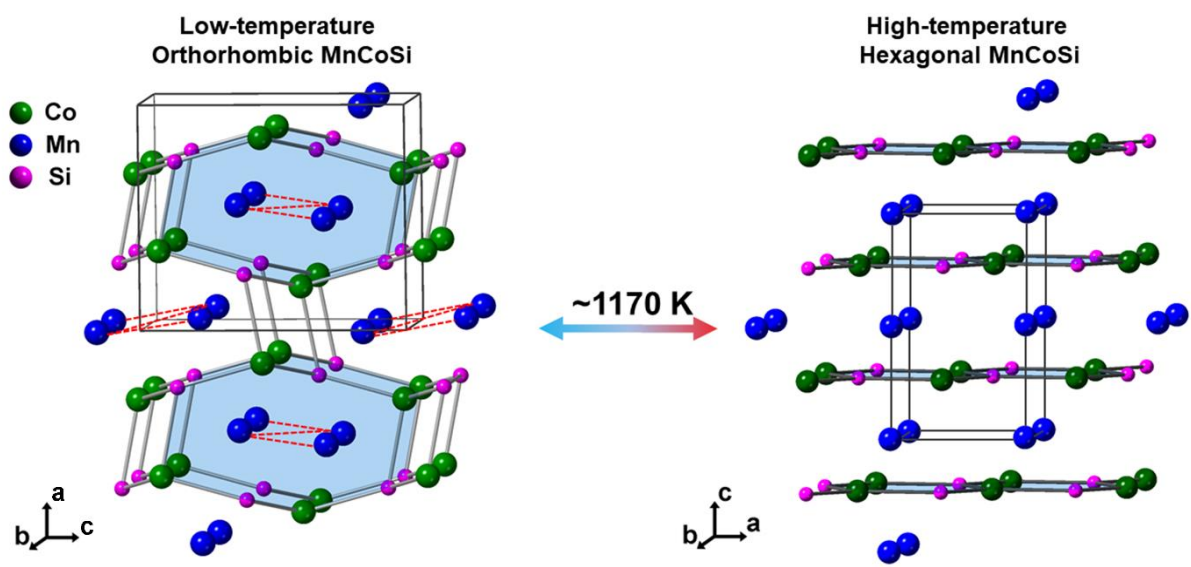

**Fig. 1 Crystalline structures of MnCoSi.** The low-temperature orthorhombic and high-temperature hexagonal crystalline structure of MnCoSi. Black solid lines refer to the unit-cell of two structures.

## Materials and Methods

### Sample synthesis

Polycrystalline $Mn_{1-x}Ni_xCoSi$ ($x$ = 0, 0.010, 0.015, 0.017, 0.020 and 0.025) samples were prepared by arc melting the appropriate amounts of high-purity raw materials under a purified argon atmosphere three times. Then, the as-cast samples were sealed in evacuated quartz ampoules and annealed at 1123 K for 60 h before slowly cooling to room temperature over 72 h. The slow-cooling treatment guaranteed magnetic homogeneity.[24]

### Magnetization, XRD and NPD characterization

The magnetic properties were characterized by a superconducting quantum interference device (SQUID, Quantum Design MPMS XL7) with the reciprocating sample option. The temperature dependence of the powder XRD (Rigaku, Smartlab) was collected using a low-temperature chamber. For each measurement at a specified temperature, the powder sample was maintained for 20 min to reach heat equilibrium. In situ variable-temperature NPD measurements ($\lambda$=1.622 Å) in the heating process were carried out on the Wombat beamline at the OPAL facility of the Australian Nuclear Science and Technology Organization (ANSTO). Structural refinements of the XRD and NPD patterns were performed using the Rietveld refinement method, and the irreducible representation analysis of the magnetic structure was carried out using the BASIREPS programme, both implemented in the FULLPROF package.[25,26]

### Lorentz TEM measurement

The thin plates for Lorentz TEM observations were prepared by focused ion beam. The temperature dependence of the magnetic domain structures was observed in a JEOL-dedicated

Lorentz TEM (JEOL2100F) equipped with liquid-nitrogen holders. To determine the in-plane spin distribution of the magnetic texture, three sets of images with under-, over-, and just (or zero) focal lengths were recorded by a charge coupled device (CCD) camera and then the high-resolution in-plane magnetization distribution map was obtained using commercial software QPt on the basis of the transport-of-intensity equation (TIE) equation. The orientation of the in-plane magnetization was depicted based on the color wheel. The crystalline orientation for the thin plate was checked by selected-area electron diffraction (SAED).

**Results**

According to the atomic occupancy rules in MnCoSi alloy,[23] when Ni atom nominally substitutes Mn atom, the Ni with more valence electrons preferably occupies Co site and then partial Co atom occupies Mn site. Thus, the occupation formula of the $Mn_{1-x}Ni_xCoSi$ system should be written as $(Mn_{1-x}Co_x)(Co_{1-x}Ni_x)Si$. The atomic occupation can be also confirmed by NPD refinement as shown in Fig. S6. Based on the atomic occupation, the lattice parameters and unit-cell volume of the $Mn_{1-x}Ni_xCoSi$ system are obtained from the refinement of XRD patterns, as shown in Fig. 2. For further details of the refinement, see Fig. S2 and Table S1. With decreasing temperature, the lattice parameters $b$ and $c$ typically decrease, while the lattice parameter $a$ shows a NTE effect. Moreover, with the introduction of Ni, the expansion of the lattice parameter $a$ upon cooling shifts to a low temperature and changes dramatically. As a result, the effect of the shrinkage of $b$ and $c$ on the unit-cell volume is compensated for; thus, ZTE behaviour is realized in Ni-containing samples, as presented in Fig. 2d. The reliability of anisotropic thermal expansion and ZTE can be also confirmed from NPD results (Fig. S7). Notably, stoichiometric MnCoSi exhibits a linear PTE in the studied temperature range with a slight inflection at approximately 230 K. When Ni substitutes minimal Mn, the samples, such as contents of $x$ = 0.020 and 0.025 exhibit ultralow CTEs ($\alpha_l$ = 6.9×10$^{-7}$ and 1.3×10$^{-7}$ K$^{-1}$, respectively) over a wide temperature range (10-190 K and 10-170 K, respectively) after experiencing a normal PTE and a weak NTE, respectively, which is about one order smaller than the Invar alloy of Fe$_{65}$Ni$_{35}$ ($\alpha_l$ = 1.4×10$^{-6}$ K$^{-1}$). The calculated CTEs and corresponding working temperature ranges are listed in Table S2 in the Supplementary Materials.

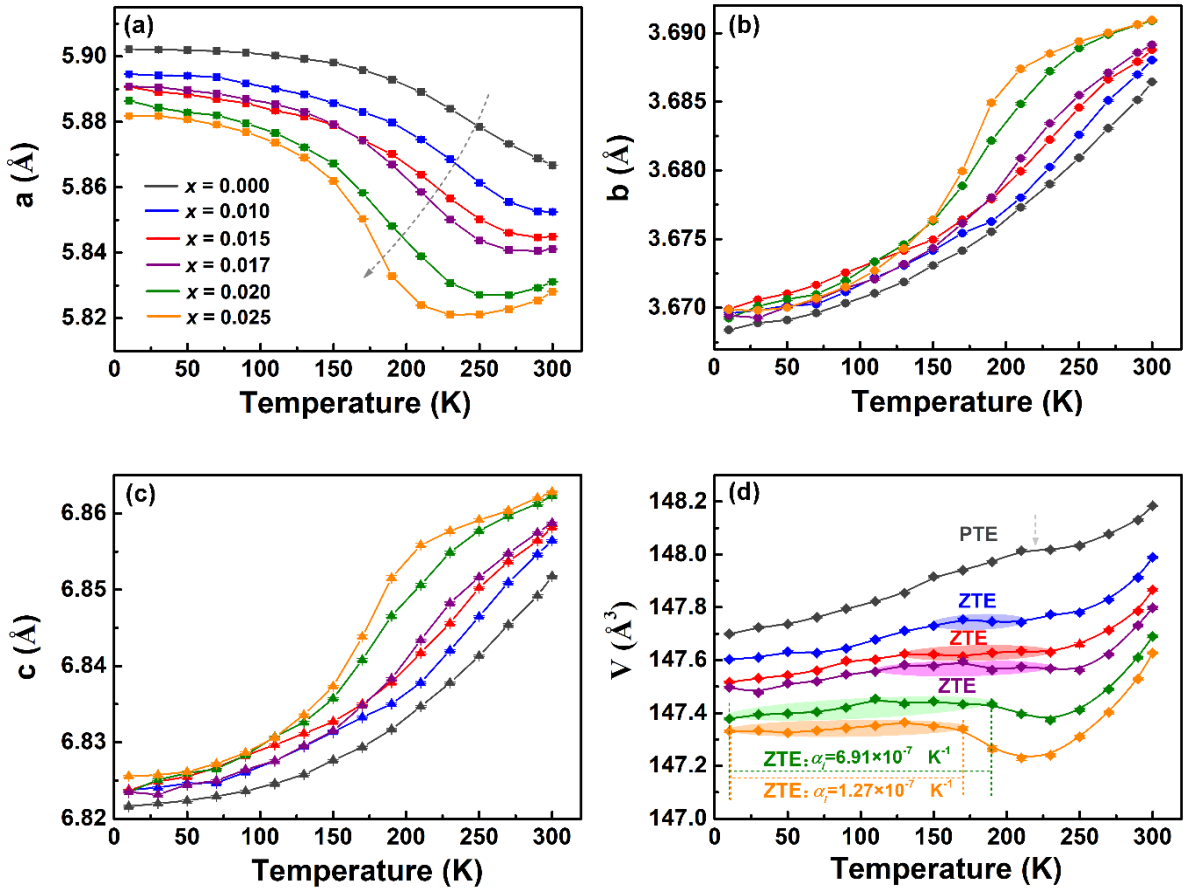

**Fig. 2 Lattice parameters and unit-cell volume of Mn$_{1-x}$Ni$_x$CoSi system.** Temperature dependence of lattice parameters **(a)** *a*, **(b)** *b*, **(c)** *c* and **(d)** unit-cell volume *V* obtained from Rietveld refinement of XRD patterns of the Mn$_{1-x}$Ni$_x$CoSi system. The linear CTEs of Mn$_{1-x}$Ni$_x$CoSi system were calculated by $\alpha_l$=1/3·d*V*/(*V$_0$*·d*T*). The ZTE working temperature range is indicated in (d).

As mentioned before, the thermal expansion properties of magnetic materials can be affected by MVE. The temperature dependence of magnetization (M-T) curve during zero-field cooling (ZFC) and field cooling (FC) processes in Fig. 3a and Fig. S3 shows that the weak magnetization of nonlinear AFM increases slowly with increasing temperature; this behaviour is interrupted by the advent of paramagnetism (PM) in stoichiometric MnCoSi alloy. Hence, the long-range magnetic order cannot be maintained above $T_0 \sim 387$ K, where $T_0$ represents the order to disorder transition temperature. It is widely reported that the introduction of external elements can effectively tune the magnetic state of TMnX (T=transition metal, X=Si or Ge) alloys.[19,27] In this work, the M-T curves of the studied $Mn_{1-x}Ni_xCoSi$ system indicate that minimal Ni additions strengthen the ferromagnetic (FM) interaction, leading to a rise in the hidden thermal-induced magnetic transition, which is similar to that of Fe-substituted MnCoSi alloys.[28] For the $x = 0.020$ sample, the AFM state smoothly transitions to the FM-like state, which is accompanied by a relatively large increase in magnetization. The magnetic transition temperature $T_t$, defined as the inflection point in the M-T curve, is presented in Table S2 in the Supplementary Materials and gradually decreases with increasing Ni content. Additionally, the establishment of FM coupling can also be examined by the magnetization behaviour. As shown by the room-temperature magnetization curves in Fig. 3b, the metastable nonlinear AFM can be easily destroyed by applying a magnetic field. A second-order and nonhysteretic metamagnetic transition displaying a sharp increase in magnetization is clearly seen in stoichiometric MnCoSi and tends to saturate at $B_{sat} \sim 3.0$ T. With increasing Ni content, $B_{sat}$ decreases (Table S2), and the field-induced metamagnetism vanishes for the $x = 0.020$ and $0.025$ samples, where only FM-like behaviour is exhibited. The macroscopic magnetic measurements indicate that the introduction of Ni can trigger a magnetic transition from the weak nonlinear AFM state to the FM-like state, where the NTE or ZTE emerges.

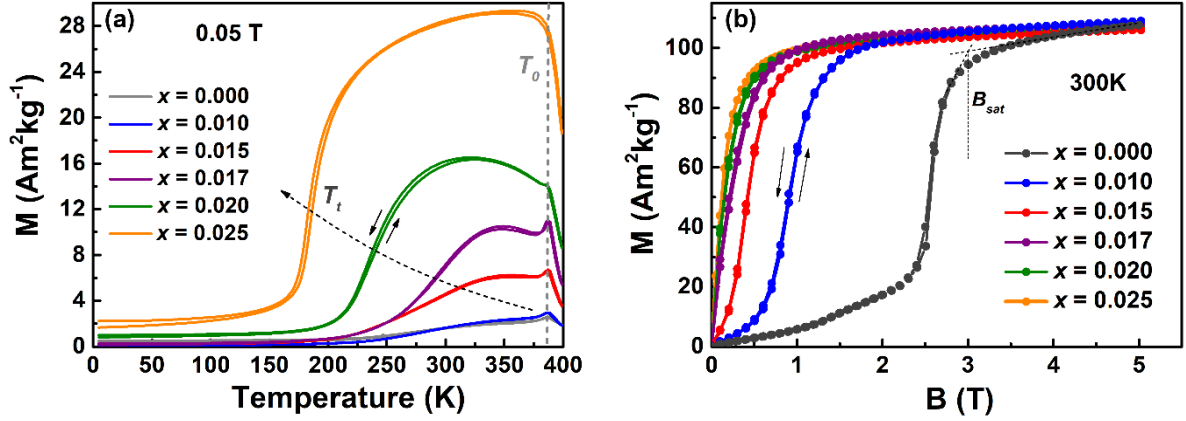

**Fig. 3 Magnetic properties of Mn$_{1-x}$Ni$_x$CoSi system. (a)** M-T curves measured at 0.05 T during the ZFC and FC processes and **(b)** magnetization curves with increasing and decreasing magnetic field at 300 K for Mn$_{1-x}$Ni$_x$CoSi system.

To further determine the evolution of the magnetic structure, the temperature-dependent NPD of the Mn$_{1-x}$Ni$_x$CoSi system was performed. As the NPD patterns of studied $x$ = 0.000, 0.015 and 0.020 samples shown in Fig. 4 and Fig. S4, with decreasing the temperature, the additional peaks at the low diffraction angles corresponding to the magnetic satellite reflections gradually appears and then splits or merges, manifesting the possible AFM order. Specifically, the refinement of isothermal NPD data at selected temperatures are also shown. At 450 K, only the nuclear scattering of orthorhombic space-group is indexed due to the sample in the disorder paramagnetic state. When the sample enters into the spin ordered state below $T_0$, the magnetic reflections can be indexed by the nonlinear magnetic structure. When the temperature further decreases to be lower than 190 K for sample $x$ = 0.020, magnetic diffraction peaks of (101)$^-$, (-101)$^-$, (202)$^-$ and (-202)$^-$ around 2θ = 18° and 40° are gradually prominent, which may be indicative of the change of nonlinear magnetic structure.

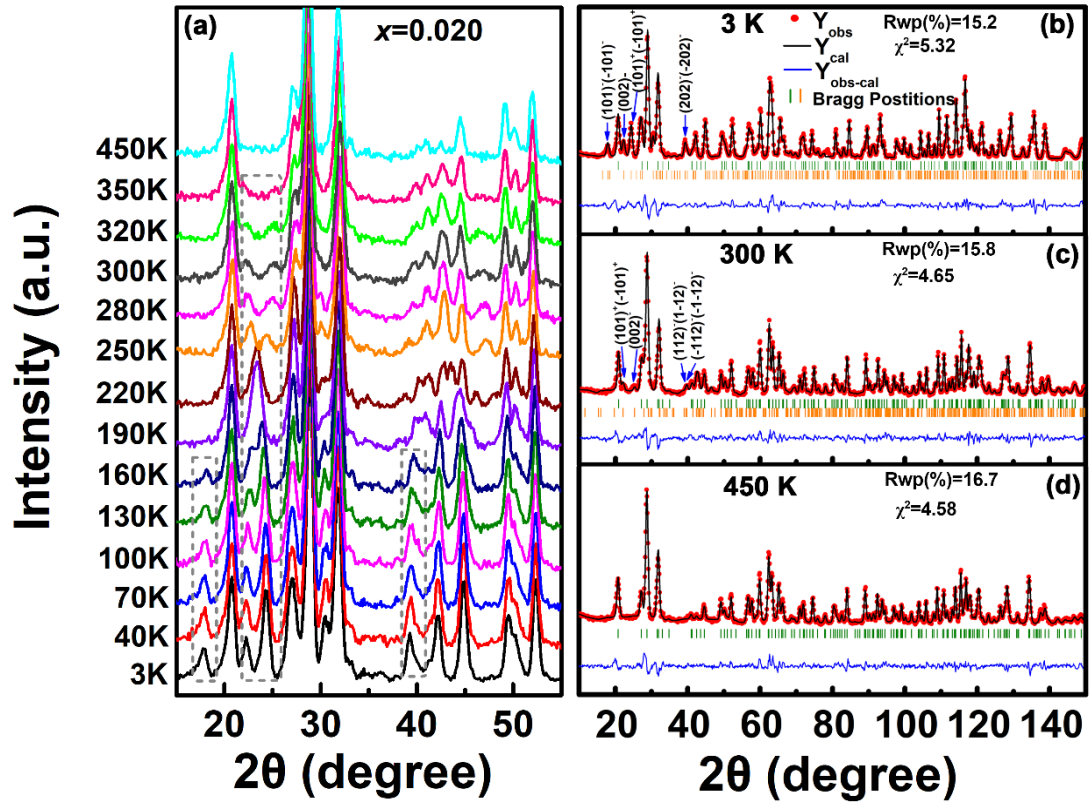

**Fig. 4 NPD patterns and the refinement results. (a)** Temperature dependence of NPD patterns for $x = 0.020$ sample. The main magnetic diffraction peaks are indicated in the grey dashed box. Specific NPD patterns at **(b)** 3 K, **(c)** 300 K and **(d)** 450 K. Experimental (red) and calculated NPD patterns (black) and their difference profiles (blue) are shown. Vertical lines indicate the peak positions for the nuclear (green) and magnetic (orange) reflection of the MnCoSi phase.

Assisted by symmetry arguments,[29] the best fit model indicates that ordered and equal moments are detected on Mn atoms (~3 $\mu_B$) or Co atoms (~0.6 $\mu_B$) (Fig. S8) at 3 K, for which the cycloidal spiral magnetic arrangement lying in *bc* plane achieves an incommensurate propagation vector $\mathbf{k} = (0, 0, k_c)$ for $x = 0.020$ sample shown in Fig. 5a-5b. This spiral magnetic structure is different from the helical magnetic structure in the literature (two NPD refinements are presented and discussed in Fig. S5),[20] which is also reported by O. Baumfeld before.[30]

Specifically, although all the atoms occupy the same crystal site (Wyckoff position *4c* (*x*, 1/4, *z*)), the wave vector group splits the magnetic Mn and Co positions into four magnetic spirals with identical ***k*** values: <Mn1, Mn3>, <Mn2, Mn4>, <Co1, Co3> and <Co2, Co4> and the magnetic spin in each cycloidal spiral rolls with a fixed angle along *c* axis. As the main carrier of the magnetic moment, two magnetic spirals of Mn atoms exhibit obvious phase differences, indicating different spin orientations in the *x* = 0.020 sample at ground state. When the temperature increases, the four groups of cycloidal AFM would transform to four groups of helical magnetic structures at about 190 K for *x* = 0.020 sample. As illustrated in Fig. 5d, the spin in each helix of the helical magnetic structure rotates in *ab* plane by a certain angle in going from layer to layer along the *c* axis. Therefore, the envelop of the projection of magnetic moments in *bc* plane is sinusoidally modulated shown in Fig. 5c. At 300 K, the two Mn helices almost rotate synchronously and show a small phase difference for *x* = 0.020 sample.

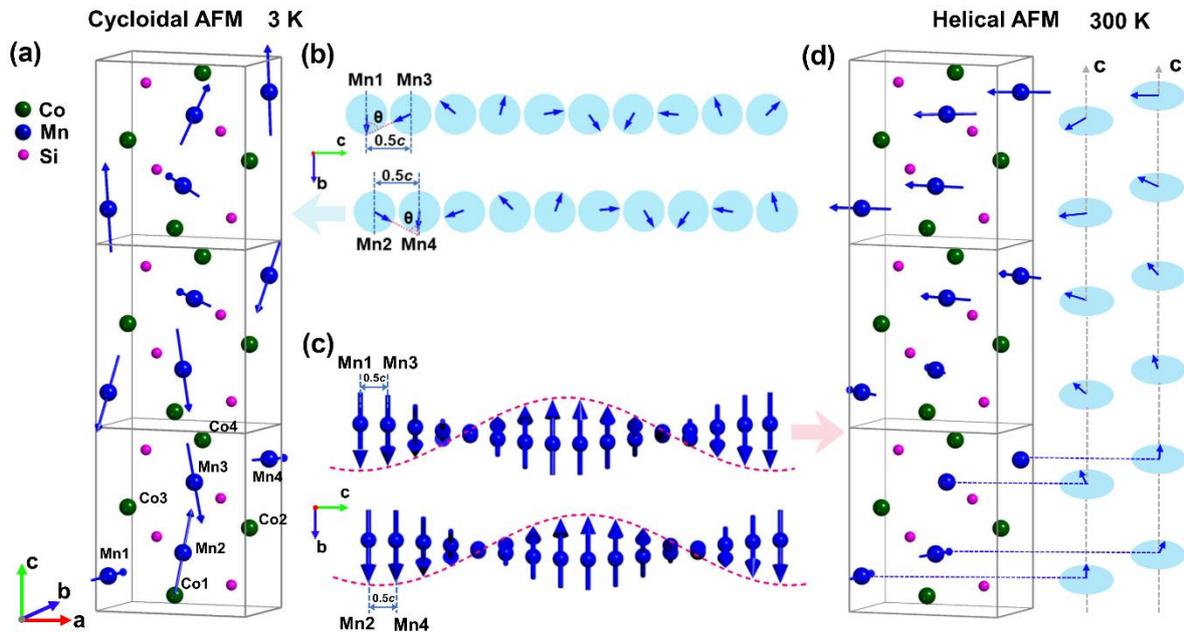

**Fig. 5 Cycloidal and helical magnetic structure of *x* = 0.020 sample. (a)** The cycloidal spiral

magnetic structure of cell $a \times b \times 3c$ at 3 K. Two groups of Mn **(b)** cycloidal (3 K) and **(c)** helical (300 K) magnetic structures viewed from the *a* axes. For clarity, Mn1 and Mn3 (Mn2 and Mn4) atoms are moved and arranged along *c* axis. **(d)** The helical magnetic structure of cell $a \times b \times 3c$ at 300 K. The moment of Co atom is smaller than the diameter of atomic symbol and is invisible in the plots.

Fig. 6 shows the related angles between different magnetic spins from the analysis of NPD refinement. Based on their atomic positions, the nearest Mn or Co atoms in each helix have *z*-coordinates that differ by 0.5, implying a phase difference of $k_c/2$ between spins and an angle between the adjacent Mn or Co atoms of $180°k_c$. With a temperature increase from 3 to 350 K, the propagation vector component $k_c$ of cycloidal or helical AFM structure for the studied samples decreases, with a relatively abrupt change near $T_t$ (Fig. 6a), indicating a decrease in the angle between the adjacent spins ($\theta_{Mn1-Mn3}/\theta_{Mn2-Mn4}$) in a cycloidal or helical magnetic chain and, correspondingly, an elongation of the magnetic spiral period. The temperature-dependent $\theta_{Mn1-Mn3}$ shows more obvious variation in Ni-containing samples, and the angle $\theta_{Mn1-Mn3}$ is further reduced (i.e., 23° at 300 K for the $x = 0.020$ sample). Moreover, the phase analysis, as shown in Fig. 6b, indicates that the angle between the nearest atoms of two Mn cycloidal or helical spirals ($\theta_{Mn1-Mn2}$) also decreases (i.e., 19° at 300 K for the $x = 0.020$ sample). Consequently, the spins of all Mn atoms tend to continuously rotate towards the parallel arrangement of *b* axis, as shown in Fig. 6c-d; thus, a FM-like spin configuration is expected in Ni-containing samples.

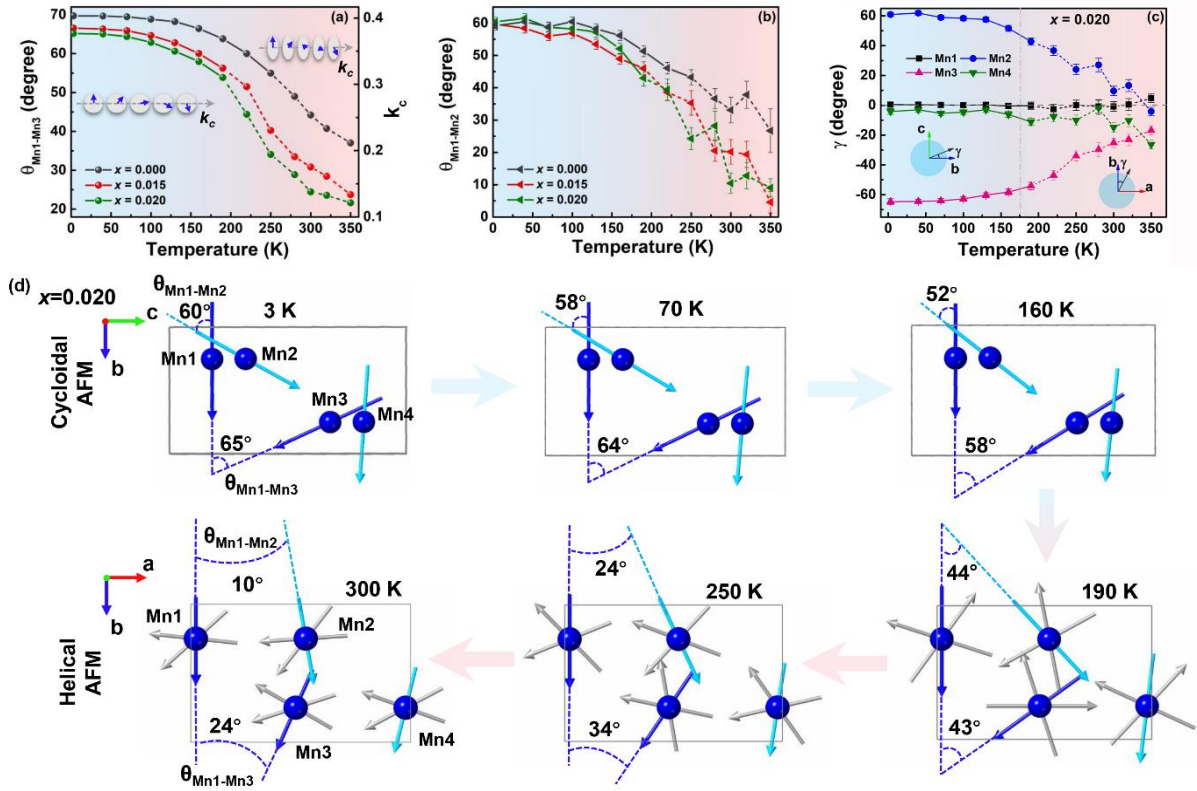

**Fig. 6 Evolution of the angles between magnetic spins.** Temperature dependence of the angle between **(a)** adjacent Mn1 and Mn3 magnetic spins ($\theta_{Mn1\text{-}Mn3}=180°k_c$), **(b)** adjacent Mn1 and Mn2 magnetic spins ($\theta_{Mn1\text{-}Mn2}$) and **(c)** Mn magnetic spins and $b$ axis ($\gamma$). The solid and dashed lines correspond to cycloidal and helical structures, respectively. Error bars are also shown in the plots; some error bars are smaller than the data symbols. **(d)** Mn cycloidal magnetic structures of cell $a\times b\times c$ for $x = 0.020$ sample viewed from the $a$ axes at 3 K, 70 K and 160 K. Mn magnetic helices of cell $a\times b\times 3c$ for $x = 0.020$ sample viewed from the $c$ axes at 190 K, 250 K and 300 K, respectively. The angles between different Mn spins are defined and given.

In addition, Lorentz TEM is also widely employed to investigate the real-space imaging of spiral magnetic structures.[31,32] We then imaged the magnetic domain structures using Lorentz TEM under zero magnetic field between 140 K and 300 K in $x = 0.020$ sample, as shown in Fig. 7 (details are shown in Fig. S9). The studied thin specimen is near the [1-10] zone axis orientation confirmed by the SAED in the inset. At 293 K, the uniform and nano-sized fine magnetic patterns with bright and dark contrast are repeatedly arranged perpendicular to the *c* axis. Based on the over- and under-focused Lorentz TEM images (Fig. S9), a TIE was adopted to characterize the spin textures of the magnetic patterns. The yellow and blue straight stripe pairs reflect the regions with opposite in-plane magnetic inductions, as indicated by the colour wheel. Together with line profile of the alternative contrast intensity, the nearly sinusoidal manner indicates the spin order is probably spiral or fan-like. It is worth mentioning that only in-plane component of moments is presented by the Lorentz TEM. Therefore, the real 3D magnetic structures of this thin MnCoSi specimen should be resolved systematically by in-situ Lorentz TEM in the future. The fine stripe-type magnetic domain can be observed during the studied temperature ranges of 140-293 K. With increasing the temperature, the width of the stripe increases. Additionally, spiral magnetic structure which contains a higher harmonic modulation of magnetic order clearly provides a pair of diffraction spots along the *c* axis close to (000) diffraction spots of SAED, as shown in the inset of Fig. 7b. By quantitatively analysis of these satellite spots, the magnetic spiral period can be calculated in Fig. 7b. When the temperature increases, the period gradually increases and accelerates at about 220 K, which is consistent with the NPD data and further validates the coherent rotation of the magnetic spins.

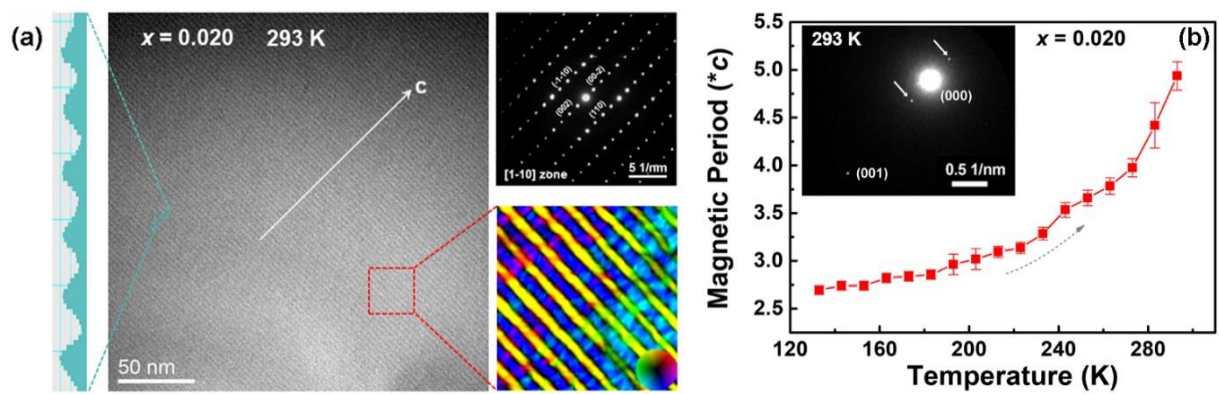

**Fig. 7 Analysis of Lorentz TEM images. (a)** The over-focused Lorentz TEM images of the domain structures at 293 K of a $x=0.020$ sample thin plate in zero magnetic field. The SAED in upper right indicates the [1-10] orientation. A line profile of the contrast intensity integrated in a dashed cyan region shown at left. The spin texture in the red dashed box obtained by TIE analysis of the Lorentz TEM data shown in lower right. **(b)** The period of the spiral magnetic structure obtained from SAED. SAED performed at camera length of 7.8 m are shown in the inset.

**Discussion**

The near-ZTE behaviour was mentioned or observed in MnCoSi-based alloys,[20,33] in which the origin of the effect was absent. The conventional mechanism underlying ZTE or NTE in magnetic alloys, such as La(Fe,Si)$_{13}$ alloys and RFe$_2$-based compounds,[11,14-16,34,35] originates from either a magnetic disorder to order transition or a large change in the magnetic moment. Here, to quantitatively uncover the contribution of magnetism to the thermal expansion behaviour of Mn$_{1-x}$Ni$_x$CoSi system, the spontaneous volume magnetostriction $\omega_m$ of sample $x$ = 0.020 is calculated by $\omega_m = \omega_{exp} - \omega_{nm}$, in which the $\omega_{exp}$ is obtained from the experimental XRD results and $\omega_{nm}$ is fitted from the nonmagnetic phase based on the Debye-Grüneisen model,[36,37] as shown in Fig. 8a. Combined with the M-T curve (Fig. 8b), the sample displays a linear PTE behaviour in the PM region. When the sample starts to enter the ordered FM-like state at below 410 K, the experimental $\omega_{exp}$ slightly deviates from the fitted $\omega_{nm}$ due to the MVE. With further decreasing the temperature, a smooth magnetic transition from FM-like state to the spiral AFM state is observed, during which the thermal expansion behaviour is significantly affected. Specifically, the negative role of magnetic $\omega_m$ gently exceeds or completely counteracts the contribution from the lattice variation, which results in a weak NTE and a wide-temperature ZTE. Moreover, it is interesting to find the $\omega_m$ and the angles between Mn spins exhibit a similar temperature dependent behaviour (Fig. 8c), which indicates an intimate relationship between the anomalous thermal expansion and spin rotation of the helical magnetic structure in MnCoSi-based alloys.

Notably, thermally-induced coherent spin rotation is also observed in stoichiometric MnCoSi. During heating, the angles $\theta_{Mn1-Mn3}$ and $\theta_{Mn1-Mn2}$ decrease from 70° and 61° to 38° and 32°, respectively. Then, the rotation is forced to cease by the disordered PM state. Therefore, this weak and partial spin rotation brings about only a small fluctuation in the thermal expansion behaviour of stoichiometric MnCoSi (shown in Fig. 2d), for which a PTE is observed over the entire temperature range. The unusual magnetic tricritical behaviour of MnCoSi results in flexible tunability of the magnetic state.[20] It is widely reported that the magnetic state of this Mn-based orthorhombic alloy with space group *Pnma* is extremely sensitive to the Mn-Mn distance $d_1$.[19,38] In this work, the introduction of Ni atoms can produce "chemical pressure" on the crystal lattice and change $d_1$. As shown in Fig. 8d, $d_1$ increases with increasing Ni content; correspondingly, nonlinear AFM tends to be FM. Therefore, the enhanced FM interaction leads to an obvious transition from the cycloidal or helical AFM state to a FM-like state, during which the magnetic spins further coherently rotate to *b* axis. Due to the robust magnetoelastic coupling,[20,33] this strong rotation gives rise to the giant spontaneous magnetostriction, particularly, the sharp contraction of *a* axis and leads to the emergence of the anomalous NTE or ZTE behaviour in the homogeneous phase. In addition, our results suggest that an appropriate internal or external stimulus, such as doping with elements, introducing vacancies or applying hydrostatic pressure or a magnetic field can strengthen the FM interaction and establish this spiral AFM-FM-type transition and then ZTE would be induced in MnCoSi-based alloys.

It is worth mentioning that the change of magnetic structures from cycloidal to helical spiral cannot be obviously revealed in the evolution of lattice parameters and magnetic properties of polycrystalline samples. While due to the intimate relations between magnetic state and Mn-Mn separation,[20] the evident step change of $d_1$ (Fig. 8d) can be observed which corresponds to the change of spiral AFM structure in MnCoSi-based alloy. Additionally, owing to the distinct easy-magnetization plane of cycloidal (*bc* plane) and helical (*ab* plane) spirals, the magnetic structures can be effectively distinguished by magnetic characterization of MnCoSi single crystal.

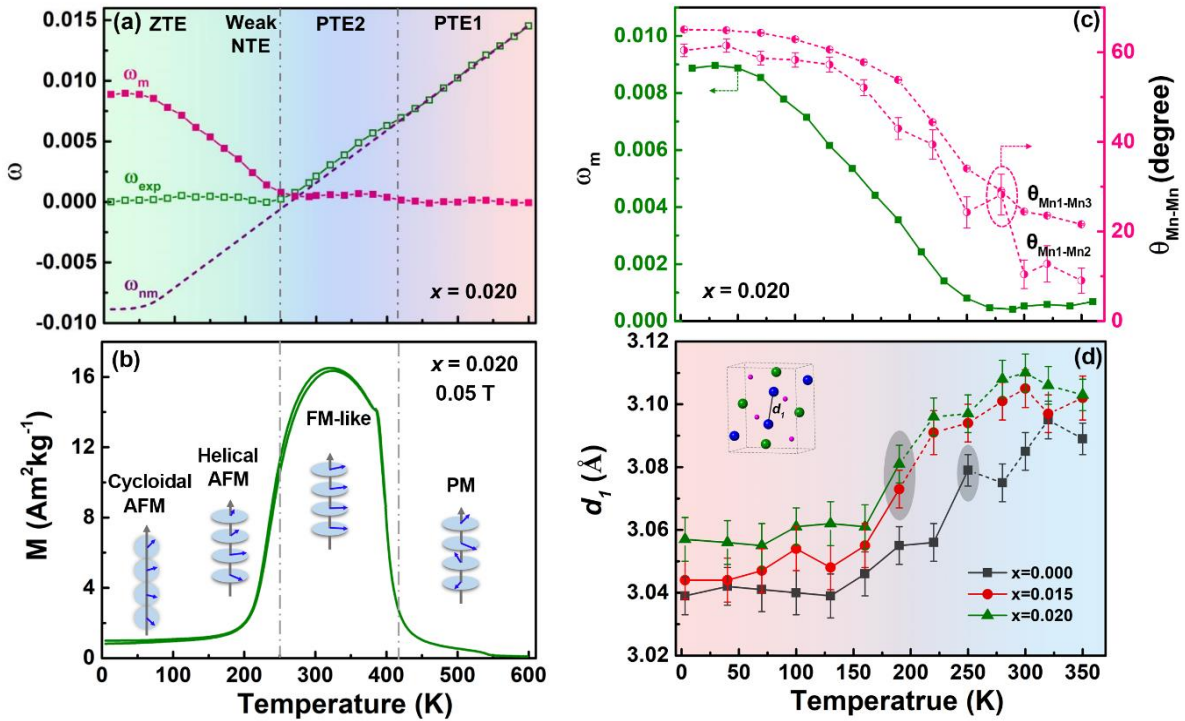

**Fig. 8 Thermal expansion properties and evolution of the magnetic structure parameters.**
**(a)** Thermal expansion of sample *x* = 0.020. The spontaneous volume magnetostriction ($\omega_m$) is calculated by subtracting the contribution of nonmagnetic part ($\omega_{nm}$) to the experimental thermal expansion data ($\omega_{exp}$). **(b)** M-T curves of sample $x$ = 0.020 at 0.05 T. **(c)** Temperature dependence of $\omega_m$ and adjacent angles between Mn spins. **(d)** The temperature dependence of Mn-Mn distance $d_1$ obtained from Rietveld refinement of NPD patterns. $d_1$ is indicated in the inset and the magnetic structure variation temperature is highlighted.

**Conclusions**

In summary, a wide-temperature ZTE effect and a new cycloidal spiral AFM structure were found in orthorhombic $Mn_{1-x}Ni_xCoSi$ alloys. Systematic magnetic measurements and in situ XRD, NPD and Lorentz TEM characterization indicated that the introduction of Ni can enhance the FM interaction and induce a transition from the spiral AFM state to a FM-like state. During this transition, the spin lying in the *bc* or *ab* plane rotates uniformly and leads to drastic changes in the lattice parameters due to magnetoelastic coupling, which results in ZTE behaviour. Moreover, this new mechanism sheds light on magnetic materials that possess this spiral AFM-FM-type transition, and ZTE or NTE materials may be discovered or designed.

**Acknowledgements**

This work was supported by the National Key R&D Program of China (2017YFA0206303), the National Natural Science Foundation of China (No. 11974406), the Strategic Priority Research Program (B) of the Chinese Academy of Sciences (CAS) (XDB33000000), and the China Postdoctoral Science Foundation (No. 2020M680735). Z.X. Cheng thanks the Australian Research Council for financial support (DP190100150).

**Conflict of interest**

The authors declare that they have no conflict of interest.

**Supplementary materials** is available for this paper.

**References**

1. Chen, J., Hu, L., Deng, J. X. & Xing, X. R. Negative thermal expansion in functional materials controllable thermal expansion by chemical modifications. *Chem. Soc. Rev.* **44**, 3522-3567 (2015).

2. Mary, T. A., Evans, J. S. O., Vogt, T. & Sleight, A. W. Negative thermal expansion from 0.3 to 1050 Kelvin in $ZrW_2O_8$. *Science* **272**, 90-92 (1996).

3. Hu, P., Chen, J., Deng, J. & Xing, X. Thermal expansion, ferroelectric and magnetic properties in $(1-x)PbTiO_3$-$x$Bi($Ni_{1/2}Ti_{1/2}$)$O_3$. *J. Am. Chem. Soc.* **132**, 1925-1928 (2010).

4. Azuma, M., *et al.* Colossal Negative Thermal Expansion in $BiNiO_3$ Induced by Intermetallic Charge Transfer. *Nat. Commun.* **2**, 347 (2011).

5. Yilmaz, S. & Dunand, D. C. Finite-element analysis of thermal expansion and thermal mismatch stresses in a Cu-60vol%$ZrW_2O_8$ composite. *Compos. Sci. Technol.* **64**, 1895-1898 (2004).

6. Miracle, D. B. Metal matrix composites-From science to technological significance. *Compos. Sci. Technol.* **65**, 2526-2540 (2005).

7. Liu, J., *et al.* Realization of zero thermal expansion in La(Fe,Si)$_{13}$-based system with high mechanical stability. *Mater. Des.* **148**, 71-77 (2018).

8. Guillaume, C. É. Recherches sur les aciers au nickel. Dilatations aux temperatures elevees; resistance electrique. *CR. Acad. Sci.* 125, 18 (1897).

9. Wada, H. & Shiga, M. Thermal expansion anomaly and Invar effect of $Mn_{1-x}Co_xB$. *J. Mag. Mag. Mater.* **104**, 1925-1926 (1992).

10. Salvador, J. R., Guo, F., Hogan, T. & Kanatzidis, M. G. Zero thermal expansion in YbGaGe due to an electronic valence transition. *Nature* **425**, 702-705 (2004).

11. Wang, W., *et al.* Zero thermal expansion in $NaZn_{13}$-type La(Fe,Si)$_{13}$ compounds. *Phys. Chem. Chem. Phys.* **17**, 2352-2356 (2015).

12. Dan, S., Mukherjee, S., Mazumdar, C. & Ranganathan, R. Zero thermal expansion with high Curie temperature in Ho$_2$Fe$_{16}$Cr alloy. *RSC Adv.* **6**, 94809-94814 (2016)..

13. Li, W., *et al.* Strong coupling of magnetism and lattice induces near-zero thermal expansion over broad temperature windows in ErFe$_{10}$V$_{2-x}$Mo$_x$ compounds. *CCS Chem.* **2**, 1009-1015 (2020).

14. Song, Y., *et al.* Zero thermal expansion in magnetic and metallic Tb(Co,Fe)$_2$ intermetallic compounds. *J. Am. Chem. Soc.* **140**, 602-605 (2018).

15. Song, Y., *et al.* Transforming thermal expansion from positive to negative: the case of cubic magnetic compounds of (Zr,Nb)Fe$_2$. *J. Phys. Chem. Lett.* **11**, 1954-1961 (2020).

16. Song, Y., *et al.* Negative thermal expansion in (Sc,Ti)Fe$_2$ induced by an unconventional magnetovolume effect. *Mater. Horiz.* **7**, 275-281 (2020).

17. Takenaka K. & Takagi, H. Giant negative thermal expansion in Ge-doped anti-perovskite manganese nitrides. *Appl. Phys. Lett.* **87**, 261902 (2005).

18. Miao, P., *et al.* Large magnetovolume effect induced by embedding ferromagnetic clusters into antiferromagnetic matrix of cobaltite perovskite. *Adv. Mater.* **1605991**, 1-6 (2017).

19. Gercsi, Z. & Sandeman, K. G. Structurally driven metamagnetism in MnP and related Pnma compounds. *Phys. Rev. B* **81**, 224426 (2010).

20. Barcza, A., Gercsi, Z., Knight, K. S. & Sandeman, K. G. Giant magnetoelastic coupling in a metallic helical metamagnet. *Phys. Rev. Lett.* **104**, 247202 (2010).

21. Sandeman, K. G., Daou, R., Özcan, S., Durrell, J. H., Mathur, N. D. & Fray D. J. Negative magnetocaloric effect from highly sensitive metamagnetism in CoMnSi$_{1-x}$Ge$_x$. *Phys. Rev. B* **74**, 224436 (2006).

22. Gong, Y. Y., *et al.* Textured, dense and giant magnetostrictive alloy from fissile polycrystal. *Acta Mater.* **98**, 113-118 (2015).

23. Liu, J., *et al.* Large, low-field and reversible magnetostrictive effect in MnCoSi-based

metamagnet at room temperature. *J. Mater. Sci. Technol.* **76**, 104-110 (2021).

24. Morrison, K., *et al.* The magnetocaloric performance in pure and mixed magnetic phase CoMnSi. *J. Phys. D: Appl. Phys.* **43**, 195001 (2010).

25. Rietveld, H. M. A profile refinement method for nuclear and magnetic structures. *J. Appl. Crystallogr.* **2**, 65-71 (1969).

26. Rodriguez-Carvajal, J. Recent advances in magnetic structure determination by neutron powder diffraction. *Physica B* **192**, 55-69 (1993).

27. Nizioł, S., Bombik, A., Bażela, W., Szytuła, A. & Fruchart D. Mgnetic phase diagram of $Co_xNi_{1-x}MnGe$. *Solid State Commun.* **42**, 79-83 (1982).

28. Morrison, K., *et al.* Measurement of the magnetocaloric properties of $CoMn_{0.95}Fe_{0.05}Si$: Large change with Fe substitution. *Phys. Rev. B* **78**, 134418 (2008).

29. Izyumov, Y. A., Naish, V. E. & Ozerov, R. P. *Neutron Diffraction of Magnetic Materials* (Consultants Bureau, Plenum Publishing Corporation, New York, 1991).

30. Baumfeld, O. L. Magnetoelastic coupling and tricritical metamagnetism. PhD dissertation, Imperial College London, 2017.

31. Uchida, M., Onose, Y., Matsui, Y. & Tokura, Y. Real-space observation of helical spin order. *Science* **311**, 359-361 (2006).

32. Togawa, Y., *et al.* Chiral magnetic soliton lattice on a chiral helimagnet. *Phys. Rev. Lett.* **108**, 107202 (2012).

33. Barcza, A., et al. Magnetoelastic coupling and competing entropy changes in substituted CoMnSi metamagnets. *Phys. Rev. B* **87**, 064410 (2013).

34. Li, S., *et al*. Zero thermal expansion achieved by an electrolytic hydriding method in $La(Fe,Si)_{13}$ compounds. *Adv. Funct. Mater.* **1604195**, 1 (2017).

35. Huang, R., *et al.* Giant negative thermal expansion in $NaZn_{13}$-type $La(Fe,Si,Co)_{13}$ compounds. *J. Am. Chem. Soc.* **135**, 11469-11472 (2013).

36. Sayetat, F., Fertey, P. & Kessler, M. An easy method for the determination of Debye temperature from thermal expansion analyses. *J. Appl. Crystallogr.* **31**, 121-127 (1998).

37. Moriya, T. & Usami, K. Magneto-volume effect and Invar phenomena in ferromagnetic metals. *Solid State Commun.* **34**, 95-99 (1980).

38. Staunton, J. B., dos Santos Dias, M., Peace, J., Gercsi, Z. & Sandeman, K. G. Tuning the metamagnetism of an antiferromagnetic metal. *Phys. Rev. B* **87**, 060404(R) (2013).